\documentclass[aps, prb, amsmath, twocolumn, longbibliography, 10pt]{revtex4-1}

\usepackage{graphicx} 
\usepackage{multibib} 
\usepackage{xcolor}
\newcites{supp}{ } 

\begin{document}

\title{A Light-Hole Quantum Well on Silicon}

\author{Simone Assali}
\thanks{These authors contributed equally to this work.}
\affiliation{Department of Engineering Physics, \'Ecole Polytechnique de Montr\'eal, C.P. 6079, Succ. Centre-Ville, Montr\'eal, Qu\'ebec, Canada H3C 3A7}
\author{Anis Attiaoui}
\thanks{These authors contributed equally to this work.}
\affiliation{Department of Engineering Physics, \'Ecole Polytechnique de Montr\'eal, C.P. 6079, Succ. Centre-Ville, Montr\'eal, Qu\'ebec, Canada H3C 3A7}
\author{Patrick Del Vecchio}
\thanks{These authors contributed equally to this work.}
\affiliation{Department of Engineering Physics, \'Ecole Polytechnique de Montr\'eal, C.P. 6079, Succ. Centre-Ville, Montr\'eal, Qu\'ebec, Canada H3C 3A7}
\author{Samik Mukherjee}
\affiliation{Department of Engineering Physics, \'Ecole Polytechnique de Montr\'eal, C.P. 6079, Succ. Centre-Ville, Montr\'eal, Qu\'ebec, Canada H3C 3A7}
\author{J\'er\^ome Nicolas}
\affiliation{Department of Engineering Physics, \'Ecole Polytechnique de Montr\'eal, C.P. 6079, Succ. Centre-Ville, Montr\'eal, Qu\'ebec, Canada H3C 3A7}
\author{Oussama Moutanabbir}
\email{oussama.moutanabbir@polymtl.ca}
\affiliation{Department of Engineering Physics, \'Ecole Polytechnique de Montr\'eal, C.P. 6079, Succ. Centre-Ville, Montr\'eal, Qu\'ebec, Canada H3C 3A7}

\begin{abstract}
The quiet quantum environment of holes in solid-state devices has been at the core of increasingly reliable architectures for quantum processors and memories.\cite{Scappucci2020,Jirovec2021,Hendrickx2021,Bulaev2005,Lawrie2020,Miyamoto2010} However, due to the lack of scalable materials to properly tailor the valence band character and its energy offsets, the precise engineering of light-hole (LH) states remains a serious obstacle toward coherent photon-spin interfaces needed for a direct mapping of the quantum information encoded in photon flying qubits to stationary spin processor.\cite{Vrijen2001,Kosaka2008,Huo2014} Herein, to alleviate this long-standing limitation we demonstrate an all-group IV low-dimensional system consisting of highly tensile strained germanium quantum well grown on silicon allowing new degrees of freedom to control and manipulate the hole states. Wafer-level, high bi-isotropic in-plane tensile strain ($>1\%$) is achieved using strain-engineered, metastable germanium-tin alloyed buffer layers yielding quantum wells with LH ground state, high $g$-factor anisotropy, and a tunable splitting of the hole subbands. The epitaxial heterostructures display sharp interfaces with sub-nanometer broadening and show room-temperature excitonic transitions that are modulated and extended to the mid-wave infrared by controlling strain and thickness. This ability to engineer quantum structures with LH selective confinement and controllable optical response enables manufacturable silicon-compatible platforms relevant to integrated quantum communication and sensing technologies. 
\end{abstract}

\maketitle

Current low-dimensional hole systems exploit predominantly heavy-hole (HH) states due to the restricted capacity of the available heterostructures.\cite{Scappucci2020,Kosaka2008} For instance, the canonical Si and Ge systems, used for decades in a plethora of devices including ultrafast transistors and qubits, rely mainly on two heterostructures consisting of Si/SiGe and Ge/SiGe quantum wells (QWs).\cite{Scappucci2020,Lawrie2020} The former yields a two-dimensional electron gas, while the latter corresponds to a two-dimensional hole gas,\cite{Scappucci2020} where the compressive strain lifts the valence band degeneracy and leaves HH states well above LH states. Interestingly, the opposite configuration, where the hole ground state is of LH character, is highly desired to control a variety of quantum processes.\cite{Vrijen2001,Huo2014,Moratis2021} In fact, a LH ground state would permit stimulated Raman transitions and coherent control of spins without an external magnetic field, provide an effective spin-photon interface, and allow fast radio-frequency control of spin, arbitrary qubit rotations through virtual excitations, and control of a magnetic impurity spin coupled to a quantum dot.\cite{Moratis2021,Sleiter2006,Reiter2011,Kosaka2009} Harnessing these processes require new material platforms to access and control LH states.\cite{Huo2014,Zhang2015,Jeannin2017} Moreover, tailoring LH quantum states in group IV semiconductors holds the promise of leveraging advanced semiconductor manufacturing for large-scale processing and integration of quantum devices. In this regard, highly tensile-strained Ge QW, where the top of the valence band is of LH type, has been a long-sought-after system toward a monolithic, wafer-level approach to control LH confinement and eventually implement Si-compatible quantum functionalities.

Early attempts to develop tensile strained Ge QWs employed In$_x$Ga$_{1-x}$As as cladding layers.\cite{Pavarelli2013,Saladukha2018} Nonetheless, this approach suffers scalability limitations and undesired cross doping associated with the mixing of Ge and III-V materials. The advent of Sn-containing group IV (Si)GeSn alloys provides an alternative to overcome these challenges by allowing an independent control of the lattice parameter and the band offsets, while being compatible with Si processing.\cite{Moutanabbir2021} Yet, the epitaxial growth of these alloys has been a daunting task due to thermodynamic constraints, which can be mitigated using out-of-equilibrium growth processes leading to device quality GeSn materials including Ge/GeSn heterostructures.\cite{Moutanabbir2021,Stange2016,Xu2019,Tai2021} In the latter Ge has been introduced exclusively as a barrier in compressively strained GeSn QWs with HH ground state due to the limited Sn content and high compressive strain in GeSn layers typically used.\cite{Stange2016,Xu2019,Tai2021} Creating highly tensile strained Ge QWs will lead to a selective confinement of holes with the possibility to engineer LH and HH states. This requires the control of GeSn growth using protocols to simultaneously achieve an enhanced Sn incorporation and a significant strain relaxation followed by the growth of Ge QW and the overgrowth of GeSn at a Sn content and strain corresponding to the targeted band offset, while keeping the interfaces atomically sharp. 

\begin{figure*}[t]
    \centering
    \includegraphics[scale=0.55]{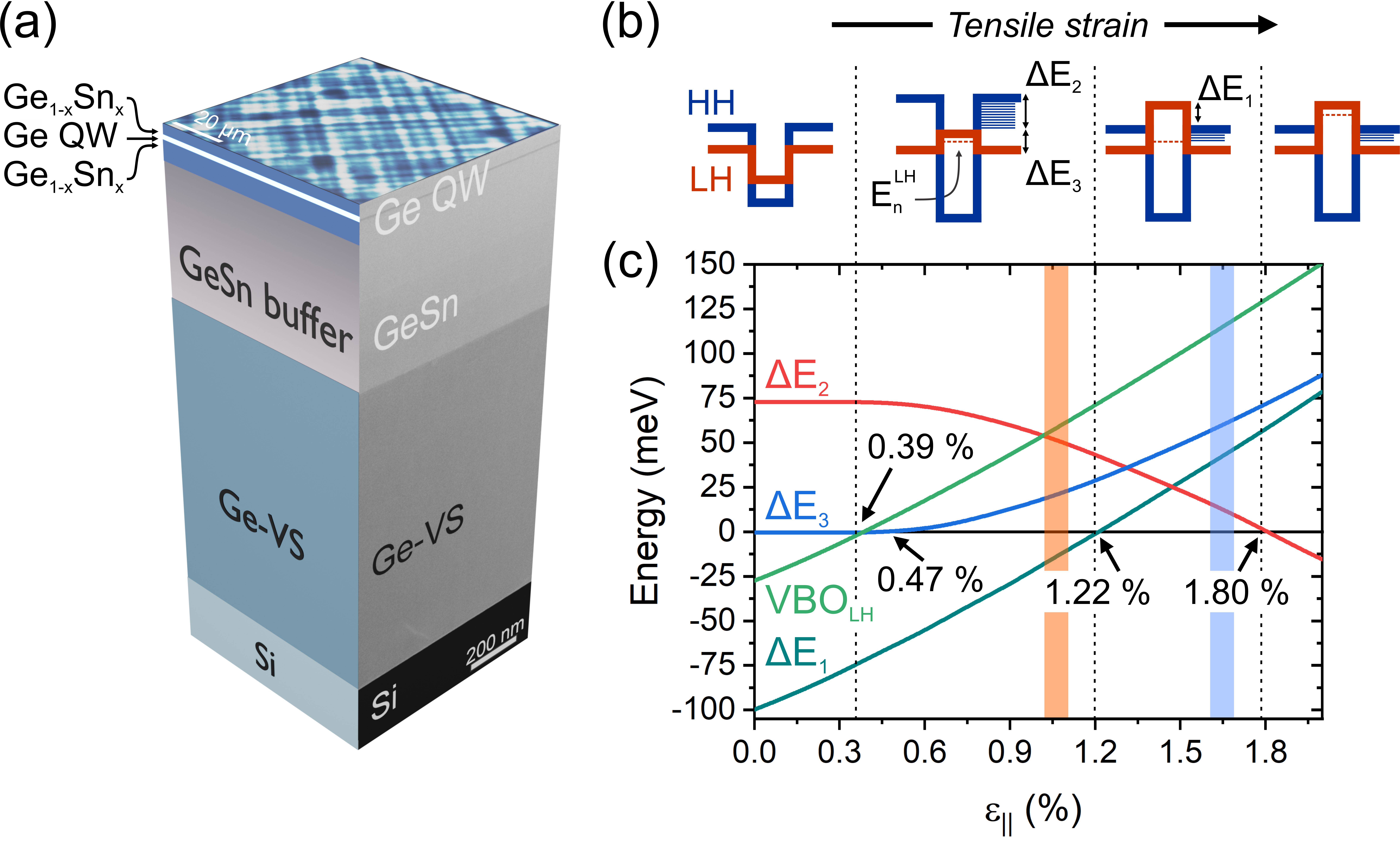}
    \caption{{\bf LH state engineering.} (a) Schematic illustration of tensile strained Ge QW heterostructure grown on Si wafer using multiple GeSn layers with increasing Sn content. A $20\,\text{$\mu$m}\times 20\,\text{$\mu$m}$ atomic force microscopy image (top) and an STEM image (side) of a typical as-grown heterostructure are superimposed on the illustration. (b) The evolution of the LH, HH band lineup as the tensile strain increases. (c) Effect of the tensile strain in Ge on the LH confinement in the QW evaluated with the 8-band $k\cdot p$.  The orange and cyan shaded rectangles represent the grown QWs at 1.1\% and 1.65\% tensile strain, respectively.} 
\end{figure*}

\medskip

\noindent {\bf Strain-engineered hole states}


Fig. 1(a) illustrates the proposed design of tensile strained Ge QWs. The integration on Si substrate is achieved through an initial Ge interlayer, commonly known as Ge virtual substrate (Ge-VS), followed by strain-engineered GeSn buffer layers with an increasing Sn content to tune the lattice parameter and control the band offsets in the heterostructure. Besides the QW thickness, the band alignment is sensitive to the thickness of GeSn cladding layers, their Sn content, and their degree of strain relaxation. Fig. 1(b) introduces the key parameters to evaluate the effects of strain on the hole states. The LH valence band offset (VBO$_\text{LH}$) is given by~: $\text{VBO}_\text{LH} = \Delta E_1 - \Delta E_2 + \Delta E_3$, where $\Delta E_1$ is the difference between LH and HH band edges, $\Delta E_2$ is the difference between the energy of the first LH subband and HH band edge, and $\Delta E_3$ is the difference between the energy of the first LH subband and the minimum LH band energy in the entire heterostructure.  In principle, LH confinement in Ge QW occurs for VBO$_\text{LH}$ higher than the thermal energy and becomes stronger as VBO$_\text{LH}$ increases. Fig.  1(c) displays the evolution of these parameters as a function of the tensile strain in the $0\text{--}2\%$ range for a $12.5\,\text{nm}$-thick Ge QW evaluated using 8-band $k\cdot p$ theory empirically optimized at $300\,\text{K}$. We note that there is no LH confinement in Ge when the tensile strain is below $0.4\%$. As the strain increases up to $1.2\%$, LH confinement is observed but in the presence of HH semi-continuum states in the cladding layer as $\Delta E_2$ is still large. $\Delta E_2$ progressively drops when strain increases, thus improving the confinement and reducing the semi-continuum HH states. These calculations highlight the role of tensile strain to fine tune the LH and HH coupling in Ge/GeSn heterostructures and indicate the need for large tensile strain ($>1\%$) for LH confinement to occur in Ge, thus setting the conditions for the epitaxial growth.

\medskip

\noindent {\bf Epitaxial growth of tensile strained Ge QW}

Ge/GeSn heterostructures were grown on Si wafers using GeH$_4$ and SnCl$_4$ as precursors according to the layout in Fig. 1(a). Fig. 2(a) exhibits a cross-sectional transmission electron microscopy (TEM) image of a typical Si-integrated heterostructure. The strain level and band offsets in Ge QW were controlled by the Sn content and the lattice parameter of the GeSn top layer (TL) and barrier layer (BR). Tuning the Sn content was achieved through a gradual adjustment of the growth temperature.\cite{Assali2019} This protocol relaxes the compressive strain in the growing GeSn layers while simultaneously enhancing the Sn incorporation. The number of growth steps defines the final Sn content and lattice parameter in the cladding layers. Fig. 2(b) displays an example of this step-graded growth, where five different GeSn buffer layers with compositions increasing from $4.9\,\text{at.}\%$ to $13.4\,\text{at.}\%$ grown at temperatures decreasing from $340\,^\circ\text{C}$ to $300\,^\circ\text{C}$. The $14.6\,\text{at.}\%$ TL and BR layers were grown at $290\,^\circ\text{C}$, while the growth of strained Ge was performed at $320\,^\circ\text{C}$. The TEM image in Fig. 2(b) shows that the gliding of misfit dislocations is promoted in the GeSn layers rather than the propagation of threading dislocations across the stacking. This results in a high crystalline quality $13.4\,\text{at.}\%$ layer (\#5) and $14.6\,\text{at.}\%$ TL. No dislocations are observed in s-Ge QW at TEM imaging scale. Fig. 2(b) also exhibits Ge and Sn  electron energy loss spectroscopy (EELS) maps confirming the Sn content in each layer. A $3.5\pm 0.1\,\text{nm}$-thick Ge QW and a $15\,\text{nm}$-thick BR layer are also visible. The uniformity of their composition is demonstrated in the close-up maps (Fig. 2(c)).

\begin{figure*}[t]
    \centering
    \includegraphics[scale=0.6]{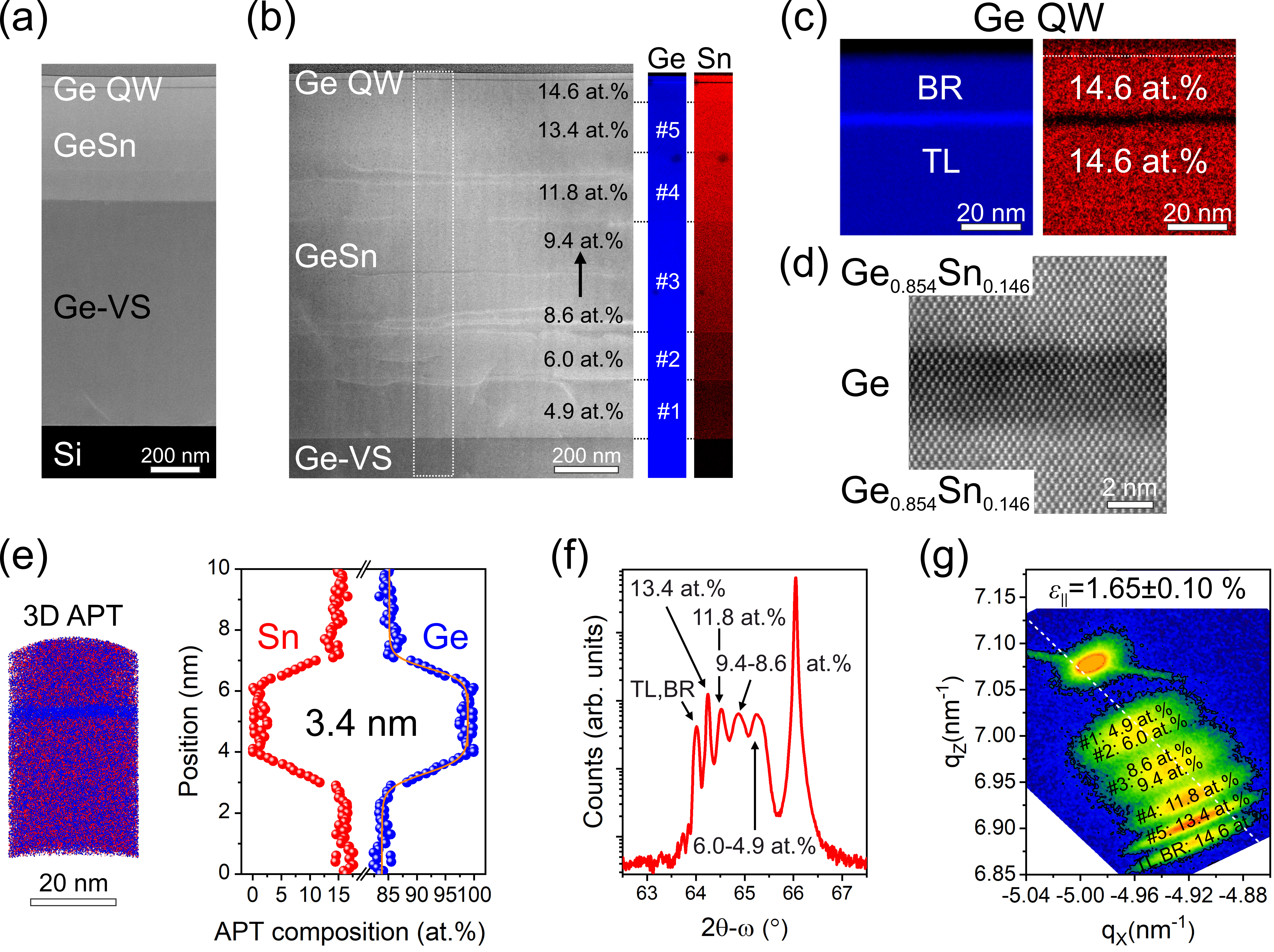}
    \caption{{\bf Structural properties of the Ge/GeSn QW heterostructure.} (a) Low-resolution STEM image of a s-Ge QW sample grown with the strain of $\varepsilon_\parallel=1.1\pm 0.1\%$. (b) TEM and EELS elemental mapping for the Ge and Sn atoms for a s-Ge QW with $\varepsilon_\parallel=1.65\pm0.10\%$. (c) High-resolution EELS map of the $3.4\,\text{nm}$ s-Ge QW. (d) HRSTEM image showing the coherent QW growth. (e) 3D APT map and compositional profile across the QW. Solid lines are the fits using Eq. \eqref{interfaces}. (f) $2\theta$-$\omega$ scan around the (004) XRD order. (g) RSM around the asymmetrical (224) reflection.}
\end{figure*}

The coherent growth is clear in the high-resolution scanning TEM (HRSTEM) image displayed in Fig. 2(d).  3D atom-by-atom mapping of the grown heterostructures was performed using atom probe tomography (APT). Fig. 2(e) shows the 3D APT map and compositional profiles of the heterostructure in Fig. 2(b), indicating the absence of Sn in the grown Ge QW and an interfacial width $w\sim 0.8\,\text{nm}$ for both interfaces. Note that regardless of the QW thickness and its strain, the top interface width of the investigated materials was always found in the $0.5\text{--}1.1\,\text{nm}$ range. However, the bottom interface broadens as the QW grow thicker. 
The (004) X-ray diffraction (XRD) spectrum (Fig. 2(f)) is characterized by sharp peaks associated with TL and BR layers are detected at angles below $64.5^\circ$. The strain in each layer was evaluated using XRD reciprocal space mapping (RSM), as exemplified in Fig. 2(g) showing the (224) RSM map of the heterostructure in Fig. 2(b). It is noticeable that GeSn buffer layers with a Sn content from $4.9\,\text{at.}\%$ (\#1) up to $11.8\,\text{at.}\%$ (\#4) exhibit a high degree of strain relaxation, while layer \#5 ($13.4\,\text{at.}\%$) and the TL ($14.6\,\text{at.}\%$) have the same $q_x$ parameter as the $11.8\,\text{at.}\%$ layer below (\#4). This observation is consistent with the high-crystalline quality of these layers (Fig. 2(b)). At $q_x^{\text{TL},\text{BR}}=4.917\,\text{nm}^{-1}$, an in-plane strain of $\varepsilon_{xx}=1.65\pm 0.10\%$ is obtained for the QW \textcolor{blue}{Section 1 in Supplementary Information, SI}. It is worth noting that the relaxation in the low Sn content $4.6\text{--}11.8\,\text{at.}\%$ buffer layers (\#1--4) prevents the compositional grading in the $13.4\,\text{at.}\%$ (\#5) layer and the $14.6\,\text{at.}\%$ (TL), thus confirming the robustness of the multi-buffer protocol to achieve stable, uniform QW growth.

\begin{figure*}[t]
    \centering
    \includegraphics[scale=0.63]{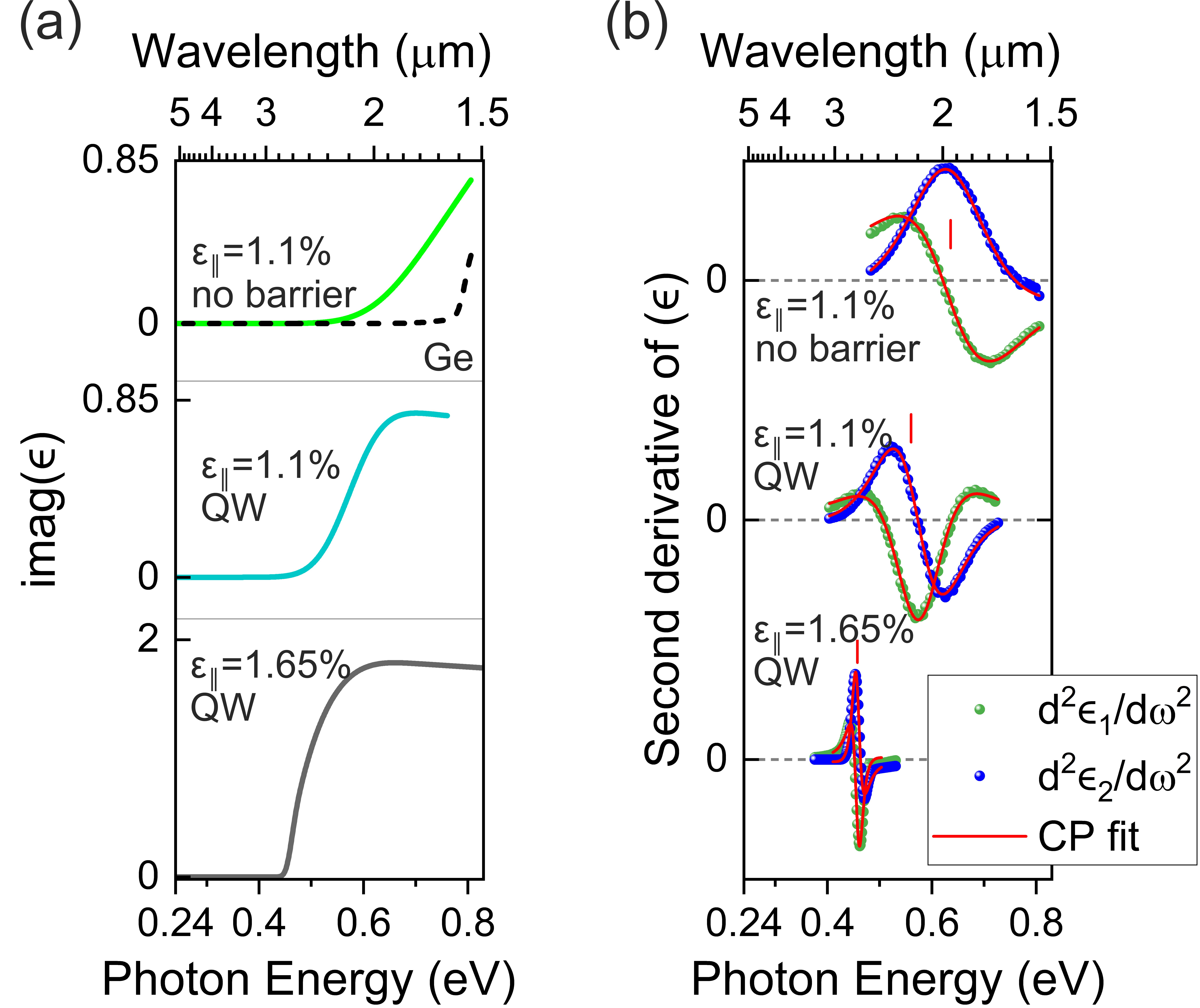}
    \caption{{\bf Effects of the quantum confinement on the optical behavior of tensile strained Ge.} (a) The imaginary dielectric function measured with IRSE for three different systems. Top, $4.5\,\text{nm}$-thick Ge layer with $\varepsilon_\parallel=1.1\pm 0.1\%$ and no GeSn barrier (green line) and bulk Ge (dashed grey line) taken from Ref. \onlinecite{Nunley2016}. Middle, $3.7\,\text{nm}$-thick Ge QW with $\varepsilon_\parallel=1.1\pm 0.1\%$ (blue cyan). Bottom, $3.5\,\text{nm}$-thick Ge QW with $\varepsilon_\parallel=1.65\pm 0.10\%$ (black line). (b) CP analysis of the corresponding dielectric function. The red vertical line indicates the CP energy position extracted after fitting simultaneously the real and imaginary part of the second derivative of the dielectric function. The fitted $d^2\epsilon_1/d\omega^2$ and $d^2\epsilon_2/d\omega^2$ are overlapped with the measured quantities as red lines.}
\end{figure*}

\medskip
\noindent {\bf Optical transitions and critical points}

To examine the electronic and optical properties of the grown heterostructures, Fourier transform infrared spectroscopy was coupled with spectroscopic ellipsometry (FTIR-SE, IRSE hereafter). Several sets of heterostructures were prepared including two sets of QWs at a tensile strain of $1.1\%$ and $1.65\%$, and uncapped tensile strained Ge layers as reference. IRSE directly measures the dielectric function and therefore is sensitive to critical points and quantization effects in the band structure. To accurately map the optical properties of tensile strained Ge, a series of reference materials were prepared for each set of heterostructures to decouple and extract the individual contributions of different layers. In this process, the absorptance measurements were combined with IRSE analyses in conjunction with 8-band $k\cdot p$ simulations. The details are outlined in \textcolor{blue}{SI (Sections 2 and 3)}. The critical point lineshape is obtained from the differentiation of the measured and simulated dielectric function ($\epsilon = \epsilon_1 +i\epsilon_2$) using the critical point parabolic band (CPPB) model.\cite{Aspnes1983} The generic standard critical-point lineshape is given by~:\cite{Lautenschlager1987_1,fujiwara2007} 

\begin{equation}\label{lineshape}
    \frac{d^2\epsilon}{d\omega^2} = \begin{cases}
        n(n-1)Ae^{i\phi}\left(\omega - E_j + i\Gamma\right)^{n-2} &\text{if $n\neq 0$} \\
        Ae^{i\phi}\left(\omega - E_j + i\Gamma\right)^{-2} &\text{if $n=0$}
    \end{cases}
\end{equation}

\noindent where $E_j$ is the threshold energy, $\Gamma$ is the broadening, and $A$ is a measure of the oscillator strength, while $\phi$ represents the amount of mixture of adjacent CPs due to excitonic effects. The exponent $n$ equals $-0.5$, $0$, $0.5$, or $-1$ for one-dimensional (1D), 2D, 3D, or excitonic interactions, respectively. 
The imaginary part of the measured dielectric function ($\epsilon_2$) of 3 sets of samples at roughly the same Ge QW thickness ($4.0\pm 0.5\,\text{nm}$) is displayed in Fig. 3. For comparison, $\epsilon_2$ of bulk Ge (black dashed line\cite{fujiwara2007}) is overlapped with that of the $1.1\%$ tensile strained uncapped layer. The CP position for the latter is located at $635.9\,\text{meV}$, whereas that of bulk Ge is at $795.4\,\text{meV}$. This difference is expected as the tensile strain induces a reduction in the direct band gap transition. $\epsilon_2$ measurements are also consistent with photoreflectance spectroscopy studies on $1.1\%$ tensile strained Ge/InGaAs.\cite{Saladukha2018} Interestingly, the measured $\epsilon_2$ of Ge QW at the same  strain (blue cyan line in Fig.3) is redshifted relative to that of uncapped layer and the corresponding CP energy difference is $75.9\,\text{meV}$. Moreover, the nature of CP changes drastically from 2D ($n=0$) Van Hove singularities in the uncapped layer to a discrete excitonic ($n=-1$) lineshape in Ge QW. Additionally, the obtained phase angle $\phi\neq 0$ corresponds to a Fano profile, meaning that the lineshape results from the interaction of the discrete excitation with a continuous background.\cite{Lautenschlager1987_2} It is noteworthy that, in addition to the CP energy shift, the broadening $\Gamma$ decreases slightly from $151\,\text{meV}$ in the uncapped layer to $114\,\text{meV}$ in QW. This observed change in CP lineshape and energy indicates that the quantum confinement, narrows the Ge optical transition. By increasing the strain from $1.1\%$ to $1.65\%$, the QW CP energy is reduced by $102.5\,\text{meV}$ and the broadening $\Gamma$ by 8-fold (Fig. 3). As discussed below, $\Gamma$ narrowing is indicative of a more spatially localized LH wavefunction in the QW.

\begin{figure*}[t]
    \centering
    \includegraphics[scale=0.68]{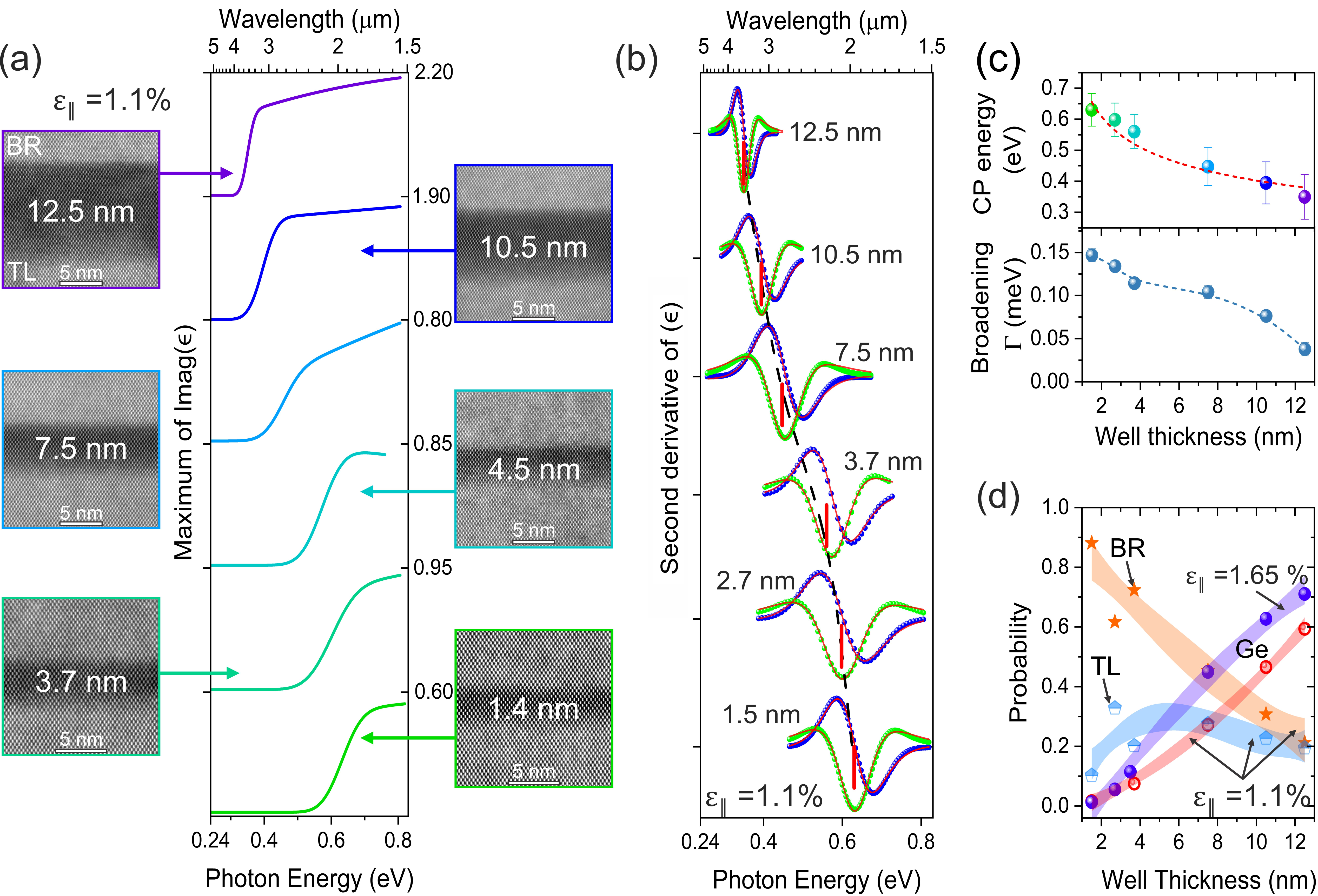}
    \caption{{\bf LH confinement in Ge QW at a variable thickness.} (a) HRTEM micrograph and imaginary part of the dielectric function acquired on the $\varepsilon_\parallel=1.1\%$ Ge QWs at a thickness in the $1.5\text{–}12.5\,\text{nm}$ range. The decrease in $\epsilon_2$ intensity as the QW thickness shrinks hints to the dielectric suppression, which is a signature of quantum confinement. (b) The $d^2\epsilon_1/d\omega^2$ (green circles) and $d^2\epsilon_2/d\omega^2$ (blue circles) and corresponding lineshape fit (red line) using Eq. \eqref{lineshape} for all samples. The red vertical lines indicate the energy position of the CP, and the dashed black line is a guide to the eye. (c) CP energy and the broadening parameter $\Gamma$ extracted from the fit in (b) plotted as a function of the QW thickness. The red line is a fit using the $a/t^b$ law for a simple, abrupt QW where $t$ is the thickness of the well and $a$ and $b=0.25\pm 0.03$ are the fitting parameters. (d) LH occupation probability in Ge or GeSn layers of the QW calculated based on 8-band $k\cdot p$ for two different tensile strains ($1.1\%$ and $1.65\%$). Increasing the tensile strain and the QW thickness leads to a relatively stronger hole confinement in the Ge QW.}
\end{figure*}

\medskip

\noindent {\bf Thickness and strain effects on LH confinement}

The effects of thickness on the electronic structure were also investigated. Fig.4(a) shows the HRSTEM images of all the studied QWs with variable thickness decreasing from $12.5\,\text{nm}$ to $1.5\,\text{nm}$ at a fixed tensile strain of $1.1\%$. Fig. 4(b) displays the evolution of $\epsilon_2$ and CP energy as a function of the thickness. We note a clear enhancement from $0.6$ to $2.2$ in the $\epsilon_2$ maximum intensity as the QW thickness increases from $1.5\,\text{nm}$ to $12.5\,\text{nm}$. This behavior is consistent with the effect of the dimension on the dielectric function.\cite{Ranjan2001} To better quantify the confinement as a function of thickness, the CP lineshape energy and broadening are displayed in Fig. 4(c). Most noticeably, the broadening increases as the thickness shrinks, while the CP energy shifts to higher energies. It has been shown in AlAs/GaAs QW that CP broadening is caused by the spread of the carrier wavefunction into the surrounding material.\cite{Weimar1996} Conversely, the observed narrowing of the CP lineshape is attributed to the confined LH wavefunction localization in the QW. As a matter of a fact, 8-band $k\cdot p$ calculations indicate that for QWs below $6\,\text{nm}$, the LH state is located $5\,\text{meV}$ below the barrier LH band leading to a spread of the wavefunction into BR and TL layers. To test this assertion, the probability to locate LH in a specific layer is evaluated using the experimental data as input in the equation $P=\sum_{n=1}^8{\int_{z_a}^{z_b}{|\phi_n(z)|^2 dz}}$, where $\phi_n(z)$ is the envelope function of the $n$th spinor component, $z_a$ and $z_b$ define the boundaries of a given layer. The obtained results demonstrate that, for a thickness above $6\,\text{nm}$, the LH state is predominantly confined in the QW and the corresponding wavefunction is more localized within the QW (Fig. 4(d)). A substantial increase in the LH localization probability in Ge is also observed at higher tensile strain.

Increasing the tensile strain in Ge QW impacts the LH confined states by reducing the LH-HH coupling and the HH continuum in the GeSn barrier as the quantization is negligible for HH. The strain also increases the LH-HH energy splitting. Although at zero in-plane wavevector pure LH states are expected, the LH-HH mixing can be significant if the LH ground state lives inside the HH-continuum. However, unlike compressively strained Ge/SiGe QWs, this LH-HH mixing can be drastically reduced in Ge/GeSn QWs by lifting the LH ground state well above the HH band edge. Furthermore, the energy gap between LH$1$ and the continuum can also be tuned by changing the QW thickness, as discussed in \textcolor{blue}{SI (Section 4.2)}.

Additionally,  LH and HH share a total angular momentum $j=3/2$ and are both subject to higher spin-orbit coupling effects compared to conduction band electrons. But, they exhibit different spin behavior when subjected to external perturbations. For instance, in systems where the confining potential lacks inversion symmetry ({\it e.g.}, under an external electric field), the spin degeneracy of LH subbands is lifted at a rate proportional to $k$ as opposed to $k^3$ for HHs. This leads to a larger LH spin splitting at a very low hole density. A LH spin also behaves quite differently under an external magnetic field compared to a HH spin. The later tends to have a very large (very small) $g$-factor in an out-of-plane (in-plane) field configuration, while for LHs it is generally the opposite \textcolor{blue}{(Fig. S11)}. Moreover, the $g$-factor of LH states in tensile strained Ge/GeSn QWs can be highly anisotropic depending on the QW thickness and strain \textcolor{blue}{(Fig. S11)}. For a tensile strain of $1.1\%$, $1.65\%$, or even $2\%$, the in-plane $g$-factor magnitude increases strongly with the well thickness, starting at $1$ for a thickness of $1.5\,\text{nm}$ and reaching $10$ at $12.5\,\text{nm}$. The out-of-plane component is also close to $1$ for thin wells, but in contrast to the in-plane component, its behavior for thicker QWs is more sensitive to strain  \textcolor{blue}{(Fig. S11)}. We exploited this high $g$-factor anisotropy in design and simulation of a range of realistic LH-spin qubit device parameters. The qubits consist of electrostatically defined quantum dots formed in Ge/GeSn QW, wherein qubit gate operations are based on electric-dipole spin resonance (EDSR). We found that the large out-of-plan $g$-factor in LH-spin qubit enables electrically driven spin flips via EDSR yielding single-qubit gate frequencies that are 2 to 3 orders of magnitude faster than reported frequencies for HH-spin qubits,\cite{Terrazos2021} thus hinting at a more robust operation of LH qubits in highly tensile strained Ge/GeSn .

\medskip

\noindent {\bf Conclusion}

We demonstrated an all-group IV LH QW on Si wafer achieved through the growth of highly tensile strained Ge on relaxed, composition step-graded GeSn buffers. The multi-buffer growth yields highly crystalline heterostructures with sub-nanometer interfacial broadening. This low-dimensional system allows to control LH-HH coupling and hence the hole states, where the LH selective confinement is obtained at a biaxial tensile strain exceeding $1\%$. The excitonic optical transitions were identified for these QWs and their energy is modulated and extended in the mid-wave infrared range by controlling strain and thickness. The ability to engineer LH states in groups IV semiconductors enables a unique Si-integrated platform combining the Ge large spin-orbit coupling, the hole quiet quantum environment, and the spin 1/2 of LH, in addition to the tunable bandgap directness and energy in strained-engineered Ge and  GeSn. These properties create valuable opportunities to implement new scalable quantum devices\cite{deLion2021} and potentially integrate on the same platform qubits, spin coherent photodetectors, and quantum emitters.

\bibliographystyle{naturemag}

\bigskip
\noindent {\bf METHODS}.

\noindent {\bf Epitaxial growth.} Samples were grown on 4-inch Si (100) wafers in a low-pressure chemical vapor deposition (CVD) reactor using ultra-pure H$_2$ carrier gas, $10\%$ monogermane (GeH$_4$) and tin-tetrachloride (SnCl$_4$) precursors. The two-step growth of the Ge-VS was performed in the $450\text{--}600\,^\circ\text{C}$ temperature range, followed by thermal cyclic annealing above $800\,^\circ\text{C}$. Subsequently, GeSn layers were grown below $350\,^\circ\text{C}$ using GeH$_4$ and SnCl$_4$ as precursors at a reactor pressure of $50\,\text{Torr}$ and constant H$_2$ flow. The GeSn composition was mainly controlled by the temperature change during growth, with the number of growth temperature steps that defines the final Sn content and lattice parameter in the cladding layers. In parallel, the initial molar fraction of the SnCl$_4$ precursor was also progressively reduced during each temperature step to compensate for the diminishing decomposition of the GeH$_4$ (constant flow) as the growth temperature decreases. The $1.65\%$ s-Ge QW consists of six different GeSn buffer layers that were grown at $340\,^\circ\text{C}$ (\#1, $4.9\,\text{at.}\%$), $330\,^\circ\text{C}$ (\#2, $6.0\,\text{at.}\%$), $320\,^\circ\text{C}$ (\#3, $8.6\text{--}9.4\,\text{at.}\%$), $310\,^\circ\text{C}$ (\#4, $11.8\,\text{at.}\%$), $300\,^\circ\text{C}$ (\#5, $13.4\,\text{at.}\%$), and $290\,^\circ\text{C}$ (TL, $14.6\,\text{at.}\%$). Next, the GeH$_4$ and SnCl$_4$ precursors supply was removed, and the sample heated to $320\,^\circ\text{C}$ for the growth of s-Ge layer using GeH$_4$ precursor. Additional GeSn growth (BR, $14.6\,\text{at.}\%$) on top was performed at $290\,^\circ\text{C}$ by re-introducing GeH$_4$ and SnCl$_4$ in the CVD reactor. A similar layout was used for the $1.1\%$ s-Ge uncapped samples (no BR), where four GeSn layers were grown at $340\,^\circ\text{C}$ (\#1, $4.9\,\text{at.}\%$), $330\,^\circ\text{C}$ (\#2, $6.0\,\text{at.}\%$), $320\,^\circ\text{C}$ (\#3, $8.2\,\text{at.}\%$), and $310\,^\circ\text{C}$ (TL, $10.5\,\text{at.}\%$). In the $1.1\%$ s-Ge QW samples the thickness of the Ge layer was controlled by the growth time, while keeping the GeSn multi-layer stacking identical between the different samples. In this layout, the GeSn layers were grown at $320\,^\circ\text{C}$ (\#1, $7\,\text{at.}\%$), $310\,^\circ\text{C}$ (\#2, $8.5\,\text{at.}\%$), $300\,^\circ\text{C}$ (TL, $10.5\text{--}12\,\text{at.}\%$, and BR $13\,\text{at.}\%$).

\noindent {\bf Materials characterization.} The nanotip preparation for the APT measurements was done in a Helios Dual-channel focused ion-beam (Dual-FIB) microscope, using the standard lift-out and tip sharpening technique. Before the nanotip fabrication, a $70\,\text{nm}$ thick Cr capping layer was deposited on the samples (using an electron-beam evaporator) to protect the top-most part of the samples from ion-implanted damage during the nanotip fabrication process. The field evaporation of individual atoms in the APT was assisted by focusing a picosecond pulsed UV laser ($\lambda = 355\,\text{nm}$), with a beam waist smaller than $5\,\text{$\mu$m}$, on the nanotip. During the APT run (in the region of interest), the laser pulse repetition rate, the evaporation rate (ion/pulse), and the laser pulse energy were maintained at $500\,\text{kHz}$, $1.0$, and $2.0\,\text{pJ}$, respectively. The base temperature and base pressure within the APT chamber were maintained at $30\,\text{K}$ and $3.2\times 10^{-11}\,\text{Torr}$, respectively. The 3D atom-by-atom reconstruction of the samples was done using the Integrated Visualization and Analysis Software (IVAS) package, using the base voltage profile. The QW interface abruptness is evaluated by fitting the Ge profile using the sigmoidal function~:\citesupp{Dyck2017}

\begin{equation}\label{interfaces}
    f(x) = A + \frac{B}{1 + e^{-\left(x_0\pm x\right)/\tau}},
\end{equation}

\noindent where $A$ is a vertical offset parameter (Ge content in the TL or BR) layer, $B$ is a scaling parameter (maximum Ge content), $x_0$ is the inflection point of the curve, and the sign of $x$ results in an increasing or a decreasing function. The interface width $w$ is then estimated as $4\tau$, while the QW thickness $t$ is taken as the difference between the inflection points at BR and TL interfaces ($t = x_0^\text{BR} - x_0^\text{TL}$).

TEM-EELS measurements were acquired using a FEI Titan 80-300 HB microscope. XRD measurements were acquired using a Bruker Discovery D8 model. In the $2\theta$-$\omega$ scans around the symmetric (004) XRD order a 3 bounces Ge (220) 2-crystals analyzer was positioned in front of the detector to minimize mosaicity effects. The in-plane lattice parameter $a_\parallel$ was estimated from the asymmetric (224) Reciprocal Space Mapping (RSM) according to the procedure discussed in Ref. \onlinecite{Assali2019}.

\noindent {\bf Theoretical calculations, 8-band $k\cdot p$.} The electronic band structure of holes and electrons is calculated with an 8-band $k\cdot p$ model for the two lowest conduction bands (CB) and the six top-most valence bands (light, heavy and split-off holes). Biaxial strain within the layers, was supplement to the $k\cdot p$ matrix via the Bir-Pikus Hamiltonian, neglecting any shear strain components.\citesupp{Eissfeller2011} Within a given layer, the strain tensor component perpendicular to growth direction (or the in-plane strain component) $\varepsilon_\parallel$ is given by $\varepsilon_\parallel = a/a_0 - 1$, with $a$ being the lattice constant of the layer and $a_0$ being the unstrained lattice constant of that layer. The out-of-plane component $\varepsilon_{zz}$ is derived from $\varepsilon_\parallel$ using $\varepsilon_{zz}=-2\varepsilon_\parallel c_{12}/c_{11}$, where $c_{11}$ and $c_{12}$ are the elastic constants. Real-space band alignment at the $\Gamma$ point is constructed by plotting in position space the eigenvalues of the $k\cdot p$ matrix with $k_x=k_y=k_z=0$ in each layer. In contrast, model solid theory\citesupp{VandeWalle1989} was employed for the real-space alignment at the L valley. For the quasi-2D problem, subband energies and envelope functions are extracted by setting $k_z\to -i \partial/\partial z$, with the derivative implemented by finite differences with uniform grid spacing of $0.01\,\text{nm}$ along the growth direction. In this case, all material parameters in the $k\cdot p$ matrix are replaced by position-dependent parameters. Finally, spurious modes are eliminated with a careful ordering of differential operators and by setting the $k^2$ terms in the CB subspace to zero, with a renormalized momentum matrix element to reproduce the CB effective mass.\citesupp{Eissfeller2011,Foreman1997} The implementation of the effect of the magnetic field on the band structure to evaluate the $g$-factor is also treated in the \textcolor{blue}{SI (section 4)}.

\noindent {\bf GeSn material parameterization.} The band structure parameter $A(x)$ of the Ge$_{1-x}$Sn$_x$ alloy is derived with Vegard’s law as $A(x)=(1-x)A^\text{Ge}+xA^\text{Sn}-x(1-x)b$, where $A^\text{Ge}$, $A^\text{Sn}$ and $b$ are the parameter for pure Ge, pure Sn, and the bowing parameter, respectively. Notably, the temperature-dependent band gaps of GeSn are calculated by applying Vegard’s law on the temperature-dependent band gaps of the alloy’s constituents. At the $\Gamma$ point for instance~: $E_{g\Gamma} (x,T)=(1-x) E_{g\Gamma}^\text{Ge} (T)+xE_{g\Gamma}^\text{Sn} (T)-x(1-x) b_\Gamma$, where $E_{g\Gamma}^\text{Ge} (T) = E_{g\Gamma}^{0,\text{Ge}}-\alpha_\Gamma^\text{Ge} T^2/(\beta_\Gamma^\text{Ge}+T)$. The equations are similar for Sn and the band gap at the L valley. Here, we used $b_\Gamma=2.46\,\text{eV}$ and $b_\text{L}=1.23\,\text{eV}$.\citesupp{Bertrand2019} An exception to the Vegard’s law was the Luttinger parameters $\gamma_1$, $\gamma_2$, and $\gamma_3$, which were interpolated only within the alloy fraction range $[0,0.2]$. All values in-betweens were interpolated using Vegard’s law with $x$ replaced by $x/0.2$. All the band structure parameters that we used in our calculations are listed in the \textcolor{blue}{SI (section 4)}.

\noindent {\bf Optical characterization.} Spectroscopic ellipsometry measurements were carried out at room temperature using an infrared variable angle spectroscopic ellipsometer manufactured by J. A. Woollam Co. The infrared variable angle spectroscopic ellipsometer system is based on a Bomem Fourier-transform infrared spectrometer and covers the $0.03$--$0.9\,\text{eV}$ range. Every single QW, as well as the corresponding buffer samples were measured at four angles of incidence ($72^\circ$, $74^\circ$, $76^\circ$, and $78^\circ$). A noticeable increase in the sensitivity of the SE parameters ($\Psi$ and $\Delta$) was observed near $76^\circ$, which is very close to the Brewster angle for Si and Ge. Thus, during the optical modeling, a special care was accorded to the fitting near this angle.

\noindent {\bf SE Modeling.} The dielectric function of the tensile strained Ge was extracted from the built optical model. The modelling methodology for each measured QW is discussed rigorously in the \textcolor{blue}{SI (sections 2 and 3)}. Besides, with regards to the optical model of the Ge well layer, the parametric semiconductor model (PSEMI) was used to quantify the dielectric function.\citesupp{Johs1998} It is important to mention that because of the presence of the interference fringes in the IR spectral range, it becomes cumbersome to use the “point-by-point” (also known as “wavelength-by-wavelength”) analysis approach to extract any physical optical property.\cite{fujiwara2007} To that end, the IRSE measurement for the buffer were combined with differential absorptance measurement with the integrating sphere to optically resolve the contribution of the buried layer in the buffer from the interference fringes and use the absorptance measurement to solidify the SE optical model critical point. It should be noted that the degree of divergence ($\chi^2$) between experimental and fitted data points is less than $2.5$ and the accuracy of the energy positions of the critical points obtained from the fitted curve is less than $\pm 3.0\,\text{meV}$.

\bibliographystylesupp{naturemag}

\bigskip
\noindent {\bf ACKNOWLEDGEMENTS}.
The authors thank J. Bouchard for the technical support with the CVD system and A. Kumar for the help with XRD analysis. O.M. acknowledges support from NSERC Canada (Discovery, SPG, and CRD Grants), Canada Research Chairs, Canada Foundation for Innovation, Mitacs, PRIMA Québec, and Defence Canada (Innovation for Defence Excellence and Security, IDEaS).

\medskip
\noindent {\bf AUTHOR CONTRIBUTIONS}.
S.A. carried out the epitaxial growth and coordinated the characterization work. A.A. performed the optical studies and developed the models to analyze the individual contributions of each layer in the investigated heterostructures. P.D.V. developed the theoretical framework, performed $k\cdot p$ simulations, and designed LH spin qubits. S.M. performed the APT studies. J.N. participate in the XRD characterization of the as-grown materials. O.M. led this research and wrote the manuscript with input from S.A., A.A., and P.D.V. All authors commented on the manuscript.

\medskip
\noindent {\bf AUTHOR INFORMATION}.
Correspondence and requests for materials should be addressed to~: oussama.moutanabbir@polymtl.ca

\medskip
\noindent {\bf COMPETING INTERESTS}.
S.A., A.A., P.D.V., and O.M. declare a related patent application that proposes silicon-integrated tensile strained germanium low-dimensional systems: International Patent Application: WO2020243831A1.

\medskip
\noindent {\bf DATA AVAILABILITY}.
The data that support the findings of this study are available from the corresponding author upon reasonable request.
\end{document}